\documentclass[aps, pra, onecolumn,superscriptaddress]{revtex4}

\usepackage{float}
\usepackage{amsmath, amssymb, graphicx, setspace}
\usepackage[colorlinks=true,linkcolor=blue,citecolor=blue,urlcolor=blue]{hyperref}

\begin{document}
\title{Dynamical nonlinear excitations induced by interaction quench in a two-dimensional box-trapped Bose-Einstein condensate}	
\author{Zhen-Xia Niu}
\affiliation{Department of Physics, Zhejiang Normal University, Jinhua 321004, China}
\author{Chao Gao}	
\email[]{gaochao@zjnu.edu.cn}
\affiliation{Department of Physics, Zhejiang Normal University, Jinhua 321004, China}
\affiliation{Key Laboratory of Optical Information Detection and Display Technology of Zhejiang, Zhejiang Normal University, Jinhua, 321004, China}
\date{\today }

\begin{abstract}
Manipulating nonlinear excitations, including solitons and vortices, is an essential topic in quantum many-body physics.
A new progress in this direction is a protocol proposed in [\href{https://doi.org/10.1103/PhysRevResearch.2.043256}{Phys. Rev. Res. \textbf{2}, 043256 (2020)}] to produce dark solitons in a one-dimensional atomic Bose-Einstein condensate (BEC) by quenching inter-atomic interaction.
Motivated by this work, we generalize the protocol to a two-dimensional BEC and investigate the generic scenario of its post-quench dynamics.
For an isotropic disk trap with a hard-wall boundary, we find that successive inward-moving ring dark solitons (RDSs) can be induced from the edge, and the number of  RDSs can be controlled by tuning the  ratio of the after- and before-quench interaction strength across different critical values.
The role of the quench played on the profiles of the density, phase, and sound velocity is also investigated.
Due to the snake instability, the RDSs then become vortex-antivortex pairs with peculiar dynamics
managed by the initial density and the after-quench interaction.
By tuning the geometry of the box traps,  demonstrated as polygonal ones, more subtle dynamics of solitons and vortices are enabled.
Our proposed protocol and the discovered rich dynamical effects on nonlinear excitations can be realized in near future cold-atom experiments.
\end{abstract}
\maketitle

\textbf{Keywords:} Bose-Einstein condensate, quench interaction, soliton, vortex

\textbf{PACS:} 03.75.Lm, 47.35.Fg, 47.32.C-, 52.35.Mw	
\section{INTRODUCTION}
As a major consequence of inter-atomic interaction, a Bose-Einstein Condensate (BEC) exhibits nonlinear properties reflecting on its excitations, and thus has attracted considerable interest.
Soliton and vortex are two types of fundamental excitations featuring nonlinear properties~\cite{kengne2021}.
They are both local density modulations that can be supported by global topology and thus can be stabilized in various systems~\cite{HZ21UCA}.
Essentially, their formation originates from a compromise between inter-atomic interactions and generic kinetics.
Concerning the BECs, according to the mean-field theory described by the Gross-Pitaevskii equation~(GPE)~[which is also called nonlinear Schr\"{o}dinger equation (NLSE) specifically in one-dimensional~(1D) space], solitons and vortices are stable separately in 1D and two-dimensional~(2D) space.

Due to their special properties, solitons and vortices show great potential application in quantum information and quantum computation~\cite{16advances}.
Therefore, manipulating these nonlinear excitations becomes an important topic in physics.
Thanks to the high degree of variability of parameters in the atomic BECs,
nowdays, solitons can be formed by various methods, including directly controlling the condensate density 
via creating shock waves~\cite{Zachary2001},
phase imprinting via tuning laser field~\cite{Denschlag2000,Fritsch2020},
and colliding two initially separated BEC~\cite{Weller2008}, etc.
While vortices can be formed by phase engineering  via interconversion of two components~\cite{Matthews99PRL},  stirring the condensate with a focused laser beam~\cite{Madison00PRL},  rotating the condensate with revolving laser beams~\cite{Abo01Science}, synthetic gauge field~\cite{Lin09Nature} and spin-orbit coupling~\cite{Wang10PRL}, etc.

However, for higher dimensions, solitons, especially that of the dark-type, are intrinsically unstable due to the snake instability mechanism~\cite{Zakharov1974,Donadello14PRL}.
In 2D, the instability can induce a soliton stripe to ring-shaped structure, and eventually toward vortices and vortex ring~\cite{kivshar2000self},
The dynamics of dark solitons in higher dimensions have been explored in the form of ring dark solitons (RDSs), which correspond to dark solitons in the radial direction.
Configuration of single and multiple RDSs can be constructed by using Raman imprinting technologies in multiple-component atomic BECs, which will be finally  split into ring-shaped vortex necklaces~\cite{Theocharis2003}.
Several approaches have been proposed to stabilize 2D solitons, including external potentials~\cite{Ma10PRA} and dipole-dipole interactions~\cite{Adhikari14PRA}.
Notice that these previous works focused on the dynamics and stability of solitons in 2D, where the number of solitons in the motion is not well controlled due to the unstable vibrational characteristics. The control on the number of solitons will be a focus in this work.

Notably, an interesting protocol used to prepare solitons in 1D BECs has been proposed by Halperin et al~\cite{Halperin2020}. The central idea is to quench the inter-atomic interaction, i.e.  change it at a given short moment of time.
The outcome of the interaction quench may be either solitons or Bogoliubov modes, and even shock waves~\cite{Jia21APS}.
Specifically, by setting the ratio of the after- and before-quench interaction strength as  $\eta^2$,  Halperin et al found that, if
$\eta$ is an integer, an initial ``half" black soliton localized at an edge of a box trap will decay into $\eta-1$ moving grey solitons without other excitation.
Such a method possesses a solid foundation elaborated by the inverse scattering theory~\cite{Gamayun2015}.

In this paper, we generalize the quench protocol to a 2D BEC and investigate the nonlinear excitations, including solitons and vortices, in its post-quench dynamics.
We find that successive inward-moving solitons can be induced in a box trap and the number of  solitons can be controlled by tuning the quench strength across different critical values.
We also find that vortex-antivortex pairs can be further produced due to the snake instability,
and their dynamics can be managed by the initial density and the after-quench interaction.
We further discuss the role of the geometry of the box traps on the dynamics of solitons and vortices.

This paper is organized as follows. In Sec.~\ref{model} we describe the 2D BEC system and the quench protocol. Then in Sec.~\ref{soliton} and \ref{vortex} we discuss the dynamics of the excited solitons and vortices due to the quench protocol. In Sec.~\ref{SF} we describe certain superfluid properties. And in Sec.~\ref{geometry} we investigate the trapping geometry effect on the quench dynamics.
Finally in Sec.~\ref{sec:summary} we present a summary of our results and outlook for future research.

\section{System and protocol}\label{model}
We first introduce the theoretic model to describe the dynamics of a 2D BEC and the protocol to manipulate the nonlinear excitation. The condensate is placed in a box trap, which has been achieved experimentally by implementing an intensity mask on the laser beam path~\cite{chomaz2015,Hueck2018,navon2021quantum}. In the following, we use natural unit $m=\hbar=1$, and adopt a dimensionless GPE to describe the dynamics of a 2D BEC,
 \begin{eqnarray}
	\label{eqn:GPE}
	i\frac{\partial\psi(\mathbf{r},t)}{\partial t}=\left[-\frac{1}{2}\nabla^2
	+V(\mathbf{r})+g|\psi(\mathbf{r},t)|^2\right]\psi(\mathbf{r},t).	
\end{eqnarray} 	
Here $\psi(\mathbf{r},t)$ is the many-body order parameter of the condensate, which is normalized as $\int d\mathbf{r} n(\mathbf{r},t)=\int d\mathbf{r} |\psi(\mathbf{r},t)|^2=1$, and $\mathbf{r}=(x,y)$ is the 2D space vector.
The external potential is set as box-type, i.e. $V(\mathbf{r}\in \Omega)=0$ and $V(\mathbf{r}\not\in \Omega)=\infty$, where $\Omega$ is  the box region.
In the following two sections, we will focus on the simplest case, i.e. a disk-shaped box trap with radius $R$, and then in Sec.~\ref{geometry} we will examine the geometric effect on the dynamics by taking different geometries of the box trap.
The dimensionless coupling constant $g$ is an effective two-body interaction strength in the 2D plane which can be reduced from the 3D counterpart.

The initial state of the system is prepared as the ground state of the condensate with interaction strength $g$, which can be numerically obtained by the imaginary time method with backward Euler centered finite difference \cite{bao2004computing}.
Note that, the bulk of the condensate is uniformly flat with a density $n_0$,
while close to the hard-wall boundary of the box trap, the condensate density features a dip, which touches zero within a scale of the healing length,  $\xi=1/\sqrt{gn_0}$. These features can be shown in Fig.~\ref{RGS} with $t=0$.
As explained in the 1D case, an initial ``half" black soliton, located in the boundary of the trap, serving as a ``seed" ,  is a key ingredient for the quench protocol.
A similar situation holds for the 2D case, where the dip of the initial density close to the boundary is also a ``half" black soliton. For a disk trap, it can be viewed as a half-RDS.

For clearly analyzing the dynamics, we further rescale the condensate density by $|\psi(\mathbf{r},t)|^2/n_0$ to normalize background density. In the following sections, we study the generic scenario of the condensate dynamics  and the nonlinear excitations, including solitons and vortices, that are triggered by instantaneously quenching the interaction strength.
The quench protocol can be achieved by either tuning the three-dimensional scattering length through
Feshbach resonance~\cite{Chin2010,Wouters2003} or by changing the $z$-axis confinement through a confinement-induced resonance~\cite{Petrov2000,Zhang2011}.

\section{solitons excited in a disk trap}\label{soliton}
\begin{figure*}[tbp]
	\centering
	\includegraphics [width=0.8\textwidth]{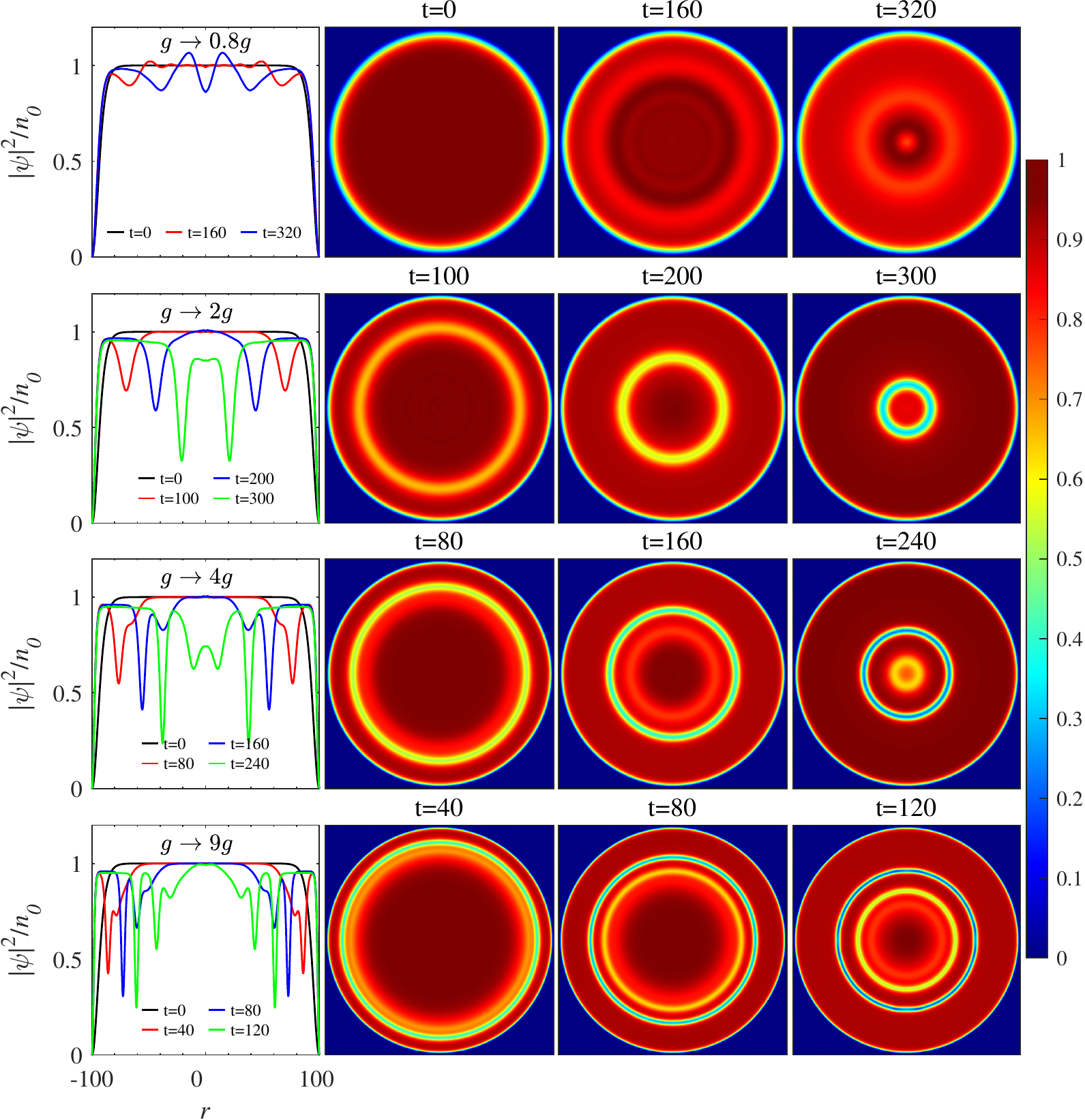}
	\caption{(Color online) Dynamical evolution of the BECs after an interaction quench in a disk trap, where solitons are created.
		First column: the density profiles of BECs along the radial direction $r$ at different moments of time after the interaction quench.
		Columns two to four: the corresponding 2D density distributions.
		The evolution of the condensates is obtained by calculating numerically Eq.~(\ref{eqn:GPE}) with the initial interaction $g=500$ and the trap radius $R=100$.}
	\label{RGS}
\end{figure*}

In this section, we investigate the excited solitons of the condensate by implementing the quench protocol in a disk-shaped box trap.
The interaction strength is quenched as  $g\rightarrow\eta^2 g$.
By a time-splitting Fourier pseudospectral method \cite{BAO2006612} to numerically solve Eq.~(\ref{eqn:GPE}) we obtain the dynamical evolution of the condensate density, see Fig.~\ref{RGS} for typical results.
We observe that the quench protocol can possibly excite moving ring grey solitons (RGSs), which can be described as moving rings of density dip below the uniform background.
According to the number of excited RGSs, we further classify the dynamics into several cases: no visible RGS (see first row in Fig.~\ref{RGS} with $g\rightarrow0.8g$), single RGS (second row with  $g\rightarrow2g$, respectively), double RGSs (third row with $g\rightarrow4g$), three RGSs (fourth row with $g\rightarrow9g$), etc.

We then analyze the detailed features of both the excited RGSs and the original half-RDS at the early stage of the dynamics.
We find that the half-RDS remains dark and does not move.
And compared to the pre-quench density profiles, the width of the half-RDS reduces if extra solitons are excited, but is fixed during the dynamics.
Moreover, stronger after-quench interaction expels more volumes $\delta V$ at the boundary, meaning narrower half-RDS,  and excites more moving RGSs.
Concerning the excited RGSs, we find that they emerge from the half-RDS at the boundary of the trap and move toward the center successively.
These excited RGSs originate from the splitting of the edge half-RDS.
Meanwhile, the later excited RGSs are shallower and faster.
While during the dynamics, the density dips of the excited RGSs gradually deep as the radius of RGSs decrease toward the center of the trap.

For a longer time, a moving RGS will shrink to the center of the trap, and then change its direction, i.e. move outward from the center and toward the edge.  If a RGS can touch the edge of the box trap, it will further be reflected by the trap edge and move inwardly again. This scenario is demonstrated in Fig.~\ref{RGS-vortex} (first row), where a single RGS is initially created by quenching interaction. In a word, the propagation of a RGS is periodic and is bounced between the trap edge and trap center. This behavior reflects the quasiparticle nature of solitons. However, such a soliton in 2D is unstable, and can be destroyed even before the touch of the box center. The instability and the transformation of solitons into vortices will be discussed in the next section.

\begin{figure}[tbp]
	\begin{center}
		\includegraphics[width=0.7\columnwidth]{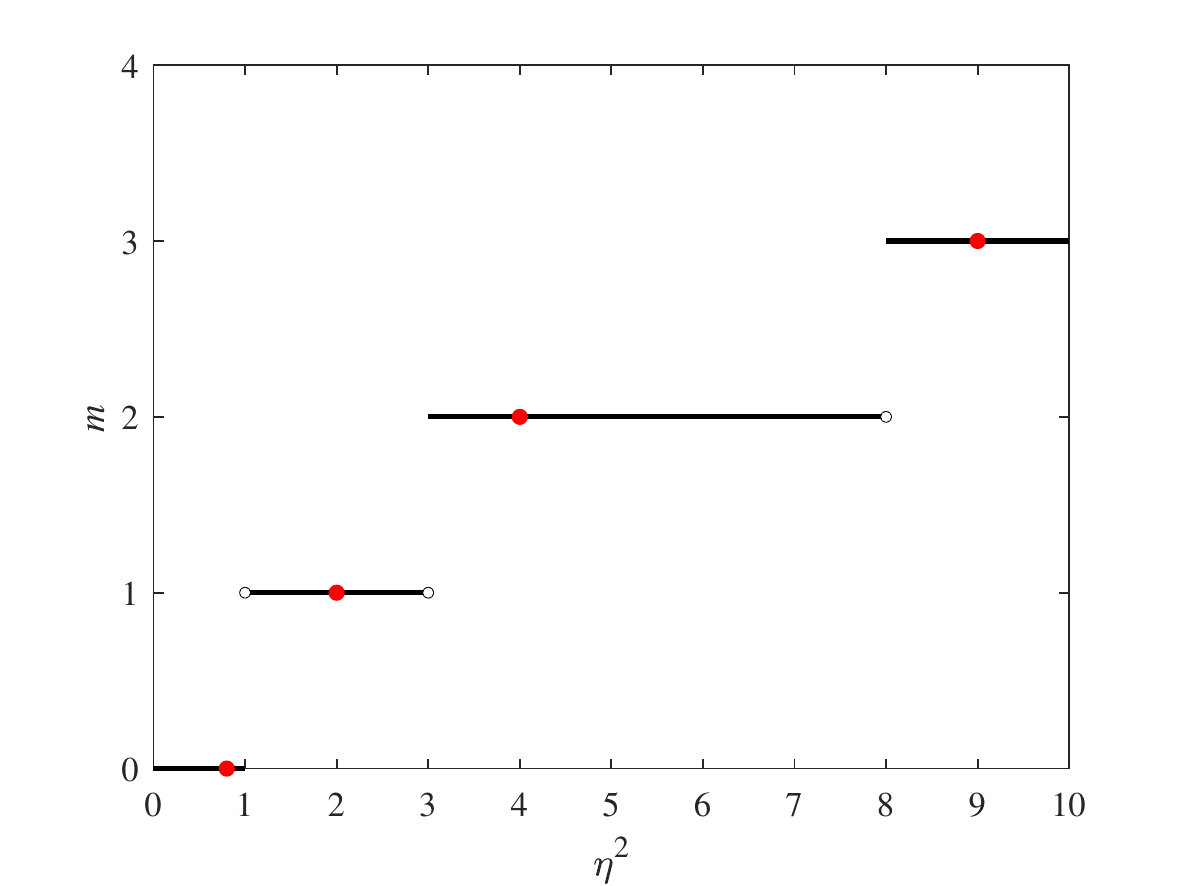}
	\end{center}
	\caption{(Color online) The number  of excited RGSs $m$ due to interaction quench $g\rightarrow\eta^2 g$ in a disk trap versus the quench  ratio $\eta^2$.
		The radius of the trap is $R=100$.
		Red dots correspond to cases shown in Fig.~\ref{RGS}.}
	\label{phase}
\end{figure}

We further discuss the condition of the number of excited RDSs.
Recalling that, in 1D BEC with uniform background and an initial black soliton under the interaction quench $g\rightarrow \eta^2 g$, exactly $2n-1$ solitons can be excited without other excitation if $\eta\equiv n$ is an integer~\cite{Gamayun2015}. These solitons include 1 black soliton remaining in the original position and  $n-1$ left-moving, $n-1$ right-moving grey solitons.
While for a 1D BEC in a box trap under the same quench,  $n-1$ grey solitons at each edge can be excited and further move away from the edges, if  $\eta\equiv n$ is an integer~\cite{Halperin2020}.
In both situations, if $\eta$ is not an integer, there will be extra excitations~\cite{Gamayun2015,Halperin2020,Jia21APS} while the number of excited solitons is the same as that of taking the ceiling of $\eta$, i.e. $n=\lceil\eta\rceil$.
For a 2D BEC, we find that the condition of the number of excited RDSs is different from that in a 1D BEC.
The dependence of the excited number of solitons $m$ versus the quench strength $\eta$  is shown in Fig.~\ref{phase}. Here the number $m$ is identified by density and phase distributions of BEC through an initial stage and before the snake instability.
We find that, when the square multiple of the initial interaction is not satisfied (e.g., $\eta=\sqrt2$ in the first row of Fig.~\ref{RGS}), an integer number of solitons are still created by the  interaction quench, while the effect of additional excitations on the newly excited RGSs is negligible.

\section{vortices excited in a disk trap}\label{vortex}
\begin{figure*}[tbp]
	\centering
	\includegraphics[width=\textwidth]{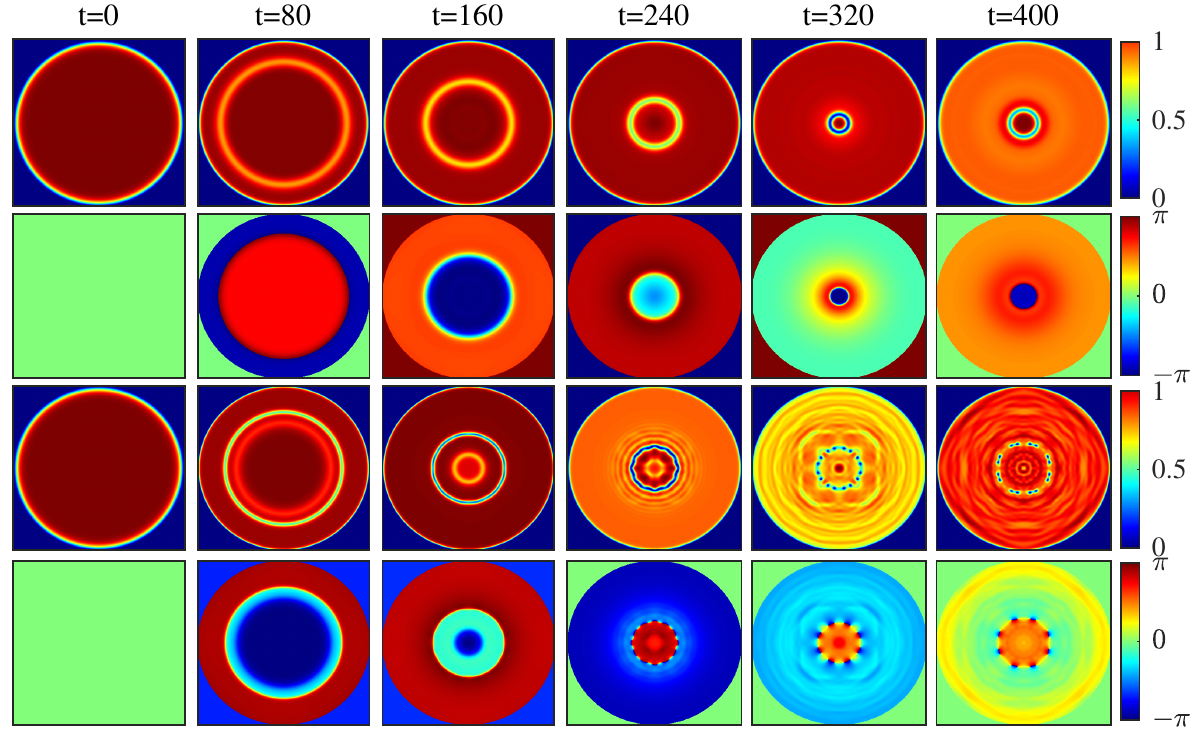}
	\caption{(Color online)  Evolutions of density and phase distributions of BECs after quenching the interaction strength $g\rightarrow1.8g$ (first and second rows) and $g\rightarrow4.5g$ (third and forth rows) in a disk trap. In the latter case, 8 vortex-antivortex pairs can be seen after $t=320$.
	The initial interaction strength and the trap radius are taken as $g=1000$ and $R=100$ respectively.}
	\label{RGS-vortex}
\end{figure*}
In this section, we investigate the vortices excited through the interaction quench in a disk trap.
We shall note that the vortices are not created in the initial stage of the evolution, but are transformed later from the moving RGSs due to the snake instability mechanism. 
In previous experimental studies, vortices were also observed as disordered decay products of dark solitons\cite{Becker2013,Donadello2014,Tamura2023}. Moreover, the effect of the symmetry in the axial direction and complex Bogoliubov-de Gennes spectrum on snake instability of RGSs have been invesigated\cite{Tamura2023,Toikka2013}.

As shown in Fig.~\ref{RGS-vortex} (second row), when the outer RGS comes across the outward moving RGS, they would decay into 8 vortex-antivortex pairs. At the same time, along with the inner RGS annihilating, irregular excitations appear in the BEC. Then these 8 vortex pairs arrange themselves on the ring moving to the boundary of the trap. Comparing to the double-RGSs created by quench interaction $g\rightarrow4g~(g=500)$ in Fig.~\ref{RGS} and $g\rightarrow4.5g~(g=1000)$ in Fig.~\ref{RGS-vortex} at $t=160$, we can find weak interaction is conducive to the formation of multiple stable RGSs.
The decay and layer structure of vortices are similar to the results of imprinted RDSs~\cite{WangWenlong2021,ZhangLiu2009,Theocharis2003}. But the interaction quench  in BEC trapped in box potentials provides a cleaner environment to observe the interaction between ring-shaped solitons, where the number of solitons can be controlled via quench strength $\eta^2$. The mismatching quench ($\eta$ is an integer in idealized quench according to the inverse scattering theory) can also excite a predetermined number of solitons, and additional excitations will not indraft other ring-shaped density wave.

\section{superfluid properties}\label{SF}
\begin{figure}[tbp]
	\begin{center}
	\includegraphics [width=0.7\columnwidth]{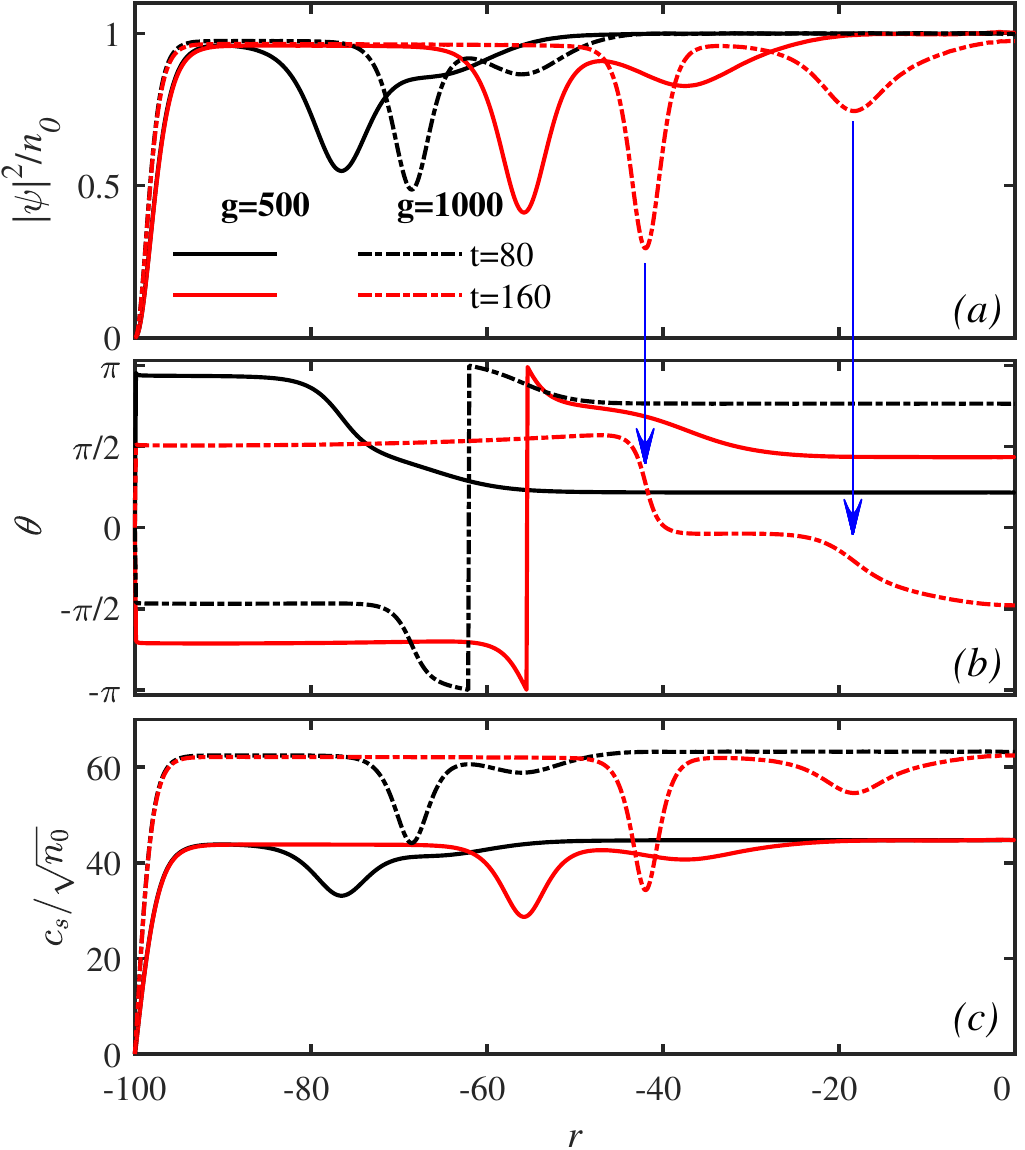}
\end{center}
\caption{(Color online) The radial distributions of the superfluid density $|\psi|^2$, phase $\theta$, and sound velocity $c_s$ of a BEC at different moments of time. Here, the initial interaction $g=500$ (solid line) and $g=1000$ (dot dashed line) and the quench $g\rightarrow4g$ are considered. The trap radius is taken as $R=100$.}
	\label{superfluid}
\end{figure}
Next, we study the superfluid properties of the condensate during its quench dynamics.
We investigate typical local quantities including the superfluid density $n(\mathbf{r},t)\equiv|\psi(\mathbf{r},t)|^2$, superfluid phase $\theta(\mathbf{r},t)\equiv\arg \psi(\mathbf{r},t)$, and sound velocity $c_s\equiv\sqrt{gn(\mathbf{r},t)}$.
In order to compare the different dynamical behaviors of the excitations, we create two moving RGSs by quenching the interaction strength $g\rightarrow4g$.
Moreover, we adopt different values of the initial interaction strength $g$ in order to investigate its role in the dynamics while fixing the trap radius.
Typical results are shown in Fig.~\ref{superfluid}.

Directly inferred from the density distribution as shown in Fig.~\ref{superfluid}(a), where dips in the radial direction correspond to ring-shaped solitons in 2D distribution,
the RGS emerging latter from the trap edge is narrower, deeper, and slower.
While stronger initial interaction would expel more volume $\delta V$ from the half-RDS at the trap edge, and after quench introduces a narrower half-RDS and deeper moving RGSs.
Along with motion of the RGSs toward the center of trap, these newly excited RGSs develop gradually deeper and narrower. Such a feature is different from that in the 1D situation~\cite{Gamayun2015}, owing to the dimensional effect.
As to the superfluid phase $\theta$, we find that the deeper solitons relate to sharper phase jumps as shown in Fig.~\ref{superfluid}(b).
Comparing to the density dips in Fig.~\ref{superfluid}(a) and sound velocity in Fig.~\ref{superfluid}(c), the stronger interaction excites faster solitons as predicted in 1D BECs.
Thus, a shallower soliton induced by  stronger interaction in the innermost density dips features a larger velocity.
However, when an innermost soliton changes moving direction at the center of the trap and approaches the edge, the instability mechanism would  induce vortex pairs.
As a result, we can decrease interatomic interaction $g$ and increase the radius $R$ of trap to prolong the time to obtain stable solitons.

\section{effect of trapping geometry}\label{geometry}
\begin{figure*}[tbp]
	\centering
	\includegraphics[width=1\textwidth]{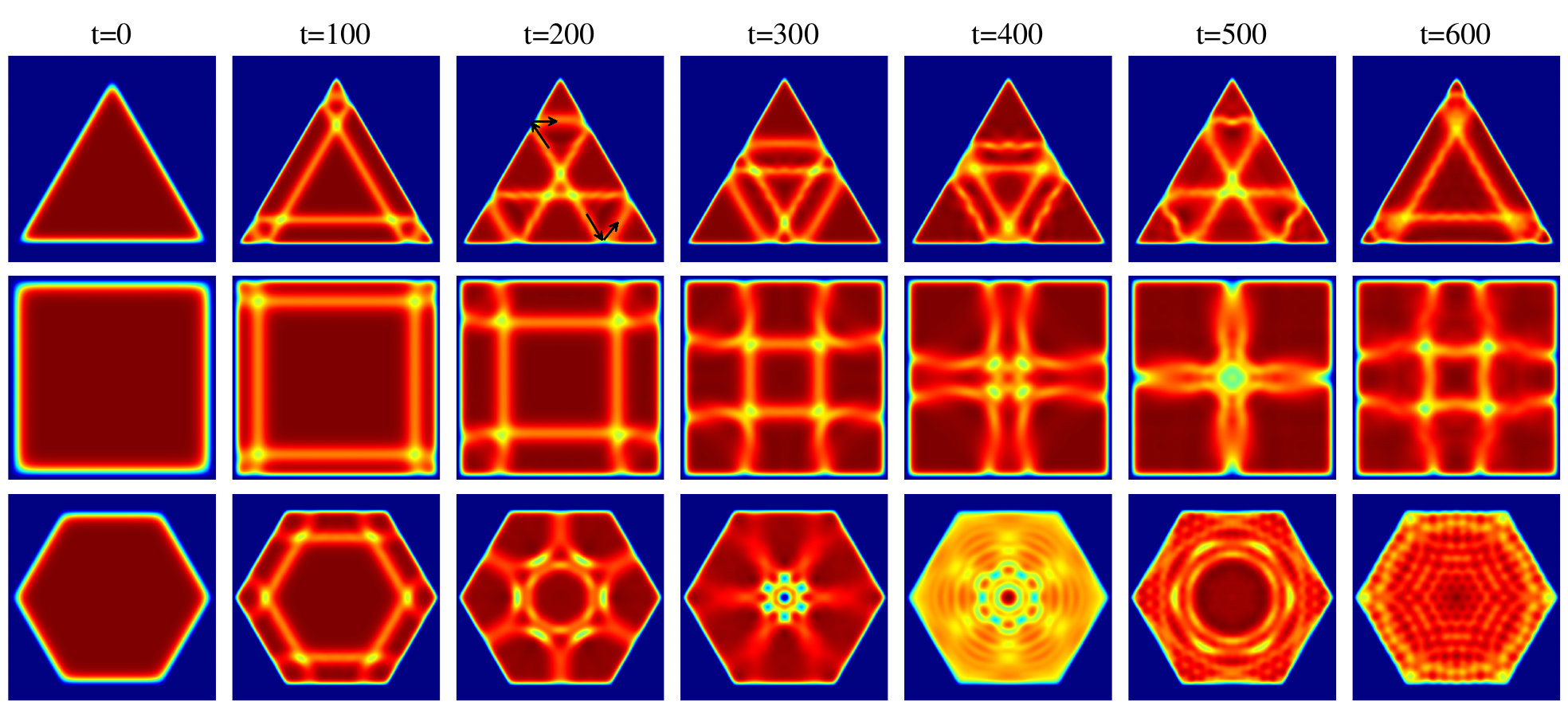}
	\caption{(Color online) Dynamical evolution of the BECs in different box traps with same interaction quench, $g\rightarrow2g$. The initial interaction strength is taken as $g=500$ and the size of the box traps is taken as $L\approx100$.}
	\label{polygon}
\end{figure*}
Finally, we investigate the effect of trapping geometry on the quench dynamics of a 2D BEC. Here,  polygonal box traps whose edges constitute regular polygons are mainly addressed, since they can be realized in cold-atom experiments nowadays~\cite{19PRX,Lv20PRL,21PRX}.  Specifically, the disk box trap can be viewed as a regular $k-$polygon one with $k\rightarrow\infty$. Typical results of the quench dynamics are shown in Fig.~\ref{polygon}, where the interaction strength is quenched as $g\rightarrow2g$, and the box shapes are taken as triangle, square, and hexagon, i.e. $k-$polygons with $k=3,4,6$.

We first note that the ground states of the condensates in all three box traps possess multiple half-black-stripe-solitons located respectively at each edge of the traps as shown in the first column of Fig.~\ref{polygon}.
After an interaction quench, a single grey-stripe-soliton can be excited from each half-black-stripe-soliton and would move in a direction perpendicular to the corresponding box edge.
Specifically for the triangle box trap, where the intersection angle $\phi$ between two adjacent edges is $\pi/3$, the emergent grey stripe solitons will be reflected on adjacent sides, which is depicted by black arrows in the first row of Fig.~\ref{polygon}.
As a result, extra stripe solitons are created from every intersection angle of the box, which further propagate in a direction perpendicular to the corresponding opposite edge.
While for cases where the intersection angle $\phi$ is larger than or equal to $\pi/2$,  the protocol  to create solitons by boundary reflection is invalid. This scenario is demonstrated by the square and hexagon box trap showed in the second and third row of Fig.~\ref{polygon}.

Moreover, the box geometry will also affect the stability of the excited grey stripe solitons. When the intersection angle of a regular polygon $\phi$ is greater than or equal to $\pi/2$, the density dips of the excited grey stripe solitons become shallower along with moving away from the edges.
And in this way, instability will induce more vortices with earlier emergence.
More complex geometries, including $k-$polygons with larger $k$, will induce even more irregular behaviors, as can be demonstrated by the case of the hexagon box trap.
Basically, this is due to the more frequent collisions between solitons and more frequent reflection by the edges.
Simply stated, the quench-induced stripe solitons in box traps with smaller intersection angles are more stable.

\section{summary and outlook}
\label{sec:summary}
In this paper, we have described a protocol to manipulate nonlinear excitations by quenching the interaction strength in 2D BECs with box traps.
Such a protocol is a generalization of that in 1D situation~\cite{Halperin2020}, while we have found several differences concerning the quench dynamics.
One is the richer dynamical behaviors in 2D, where not only solitons can be prepared, but also vortex-antivertex pairs can be induced. Moreover, the criteria to excite a certain number of solitons is different from the 1D situation.
Another one is the richer geometries of the trapping potentials that can be regulated in 2D. We have discussed their effect on nonlinear excitations.

Such a protocol can be further generalized to other systems with more complex setups.
A straightforward generalization can be done for a three-dimensional BEC where interactions can be even quenched to unitarity~\cite{eigen18Nature,gao20PRL}.
While if focused on 2D BECs, quench protocol can be applied together with peculiar dispersion, for example, that is engineered by spin-orbit coupling~\cite{Wang10PRL,Lin11Nature},
or extra special potentials including periodic ones~\cite{fleischer2003,Baifmmode2022}.
And beyond the atomic BECs, other condensates can be investigated with the quench protocol, for example, the 2D exciton-polariton system\cite{10RMP,Sun19PRB}.
All in all, given the fact that 2D configurations are ubiquitous and nonlinear excitations belong to the hot topics in the frontiers of quantum physics, we expect that this work can serve as a new starting point for manipulating various nonlinear excitations in quantum systems.

\textit{Note added--} A recent theoretic work~\cite{22CPL} studied similar quench dynamics on 2D BECs. The setup therein involves disk-shaped box traps with soft boundaries, which are different from the situation in our work. We further discuss other properties including the trapping geometry  beyond the disk type.
A  recent experimental work~\cite{PhysRevXTamura} has successfully realized ring dark solitons and vortex pairs in a 2D atomic superfluid in a circular box. While the protocol therein is different from what we proposed here.
	
\begin{acknowledgments}	
We acknowledge the useful discussion with Zhaoxin Liang and Zhe-Yu Shi.
We also thank Hikaru Tamura for  sharing their related recent work.
This work is supported by the Natural Science Foundation of Zhejiang Province, China (Grant Nos. LQ22A040006, LY21A040004, LR22A040001, LZ21A040001), and the National Natural Science Foundation of China (Grant Nos. 11835011, 12074342).
\end{acknowledgments}
\subsection*{References}
\bibliography{v2_2303_08972}
\end{document}